\documentclass[showpacs,twocolumn,preprintnumbers,amsmath,amssymb]{revtex4}
\usepackage{epsfig}
\usepackage{graphics}
\usepackage{color}

\newcommand{\etal}[1]{{\it et al.}}

\newcommand{\HHCO}{H$_2$CO}
\newcommand{\NDDD}{ND$_3$}

\begin{document}

\title{Two-dimensional trapping of dipolar molecules in time-varying electric fields}
\author{T. Junglen, T. Rieger, S.A. Rangwala, P.W.H. Pinkse, and G. Rempe}
\affiliation{Max-Planck-Institut f\"ur Quantenoptik,  Hans-Kopfermann-Str. 1,
D-85748 Garching, Germany}
\date{\today, PREPRINT}

\begin{abstract}

Simultaneous two-dimensional trapping of neutral dipolar molecules in low- and
high-field seeking states is analyzed. A trapping potential of the order of
20\,mK can be produced for molecules like \NDDD\ with time-dependent electric
fields. The analysis is in agreement with an experiment where slow molecules
with longitudinal velocities of the order of 20\,m/s are guided between four
50\,cm long rods driven by an alternating electric potential at a frequency of
a few kHz.


\end{abstract}

\maketitle

\vspace{1cm}

Cold molecules offer new perspectives, e.g. for high precision
measurements ~\cite{Hinds} and collisional physics studies
~\cite{Balakrishnan}. Pioneering work on cold molecules has been
done using cryogenic buffer gas cooling~\cite{Weinstein}. Another
promising technique for the production of cold molecules is based
on the interaction of dipolar molecules with inhomogeneous
electric fields. For example, low-field seeking molecules (LFS)
have been slowed down in suitably tailored time-varying electric
fields ~\cite{MeyerDecel} and have been trapped in inhomogeneous
electrostatic fields ~\cite{MeyerRing,MeyerTrap}. Furthermore,
efficient filtering~\cite{Bookcont} of slow LFS from an effusive
thermal source using a bent electrostatic quadrupole guide has
been demonstrated~\cite{Rangwala}.

Compared to LFS, the manipulation of high-field seeking molecules (HFS) is much
more difficult. This is mainly due to the fact that electrostatic maxima are
not allowed in free space, and hence, HFS are quickly lost on the electrodes.
Nevertheless, guiding of HFS in Keppler orbits~\cite{Loesch} and deceleration
as well as acceleration of HFS~\cite{MeyerAG} is possible. Despite this
progress in manipulating dipolar molecules, all techniques realized so far are
suited only for either LFS or HFS, not both simultaneously. However, in future
experiments with trapped samples of cold molecules, collisions or the
interaction with light fields are likely to change HFS into LFS and vice versa.
Therefore, a technique to trap both species simultaneously is vital. In this
Letter, we investigate both theoretically and experimentally a new technique,
which can trap both HFS and LFS. In particular, we report on the first
experimental demonstration of two-dimensional trapping of slow \NDDD\ molecules
from an effusive source in a bent four-wire guide driven by an alternating
electric field.

Trapping neutral particles in oscillating electric fields works as
follows~\cite{Shimizu92}. The force on a molecule in an electric field, $E$, is
given by $\vec{F}=-\vec{\nabla} W(E)$, with $W(E)$ the Stark energy of the
molecule. Polar molecules like \NDDD\ and \HHCO\ predominantly experience a
linear Stark shift, $W=s\,E$, where $s$ is the slope of the Stark shift. Other
molecules, however, and atoms usually experience a quadratic Stark effect,
$W=-\frac{1}{2}\alpha E^2$, with the polarizability $\alpha$.
\begin{figure}[htb]\begin{center}
\epsfig{file=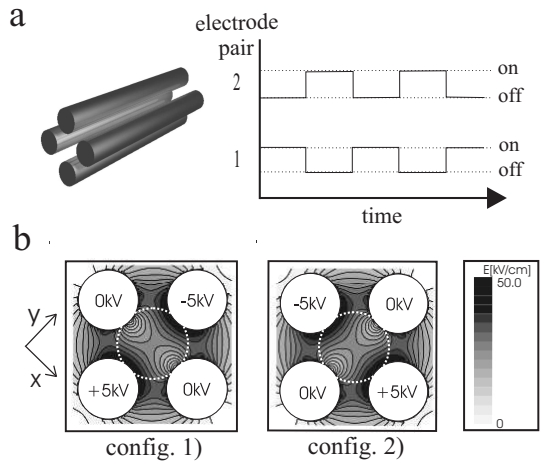,width=0.4\textwidth}%
\caption{Schematic of the four-wire setup for typical operating voltages of
$\pm5\,$kV, where $E_0=35\,$kV/cm and $\beta=2.3\times10^3\,$kV/cm$^3$. The
voltages on a pair of opposite electrodes are switched as shown in (a) in order
to produce a time-varying field that alternates between configurations 1) and
2) shown in (b) with a repetition rate in the kHz range. The radius $r_{max}$
used in simulations is indicated by the dashed circles.}%
\label{ACidea}
\end{center}\end{figure}
The inhomogeneous electric fields required for trapping can be realized in many
different ways. Our trapping configuration is sketched in Fig.~\ref{ACidea}. It
consists of 4 parallel rods with voltages as indicated. The field is rapidly
switched between two dipole-like configurations with angular frequency
$\omega$. Close to the center, the field can be expanded harmonically,
\begin{equation}
    E=E_0+H(t)\beta(x^2-y^2),
\label{TheHarmonicPotential}
\end{equation}
where $E_0$ is the field in the center and $\beta$ is (half) the curvature of
the field. The step function $H(t)=1$ if $n T<t<(n+\frac{1}{2})T$ and $H(t)=-1$
otherwise, with $T=2\pi/\omega$ the period of the driving field and $n$ an
integer.

Independent of the slope and the sign of the Stark shift, the saddle-like
dipole potential derived from Eq.(1) confines the particle in one direction and
repels it in the perpendicular direction, depending on time. Therefore, the
time average of the force is small at every position, and identical to zero for
a linear Stark shift. However, the particle performs a micromotion which is
locked to the external driving field so that the time-averaged force does not
cancel and the particle experiences a net attractive force towards the center.
This conclusion is independent from the exact form of H(t). Therefore, a
sinusoidal change between two configurations would also work. However,
instantaneous switching is more convenient to realize in the laboratory.

It is not difficult to solve the equation of motion for molecules in this
dynamic field. As the x and y degrees of freedom separate, the equation of
motion is reduced to one dimension. It reads $\ddot{x}=-H(t)\Omega^2x$, where
we have introduced $\Omega^2=2 s \beta/m$ for a linear Stark shift and
$\Omega^2=2 \alpha E_0 \beta/m$ for a quadratic Stark shift of a particle with
mass $m$. Following \cite{Morinaga94}, we search for orbits of the form
$(x(t),\dot{x}(t))=U(t) (x(0),\dot{x}(0))$ with $U(nT+t)=U(t)[U(T)]^n$.
Periodic solutions have the form $U(T)=U_2(T/2)\,U_1(T/2)$,
with %
$
    (U_1(t),U_2(t))=\left(\!\left(\!\begin{array}{cc}\cosh(t\Omega)&\frac{1}{\Omega}\sinh(t\Omega) \\
    \Omega\sinh(t\Omega)& \cosh(t\Omega)
    \end{array}\!\right)\!
,
    \left(\!\begin{array}{cc}\cos(t\Omega)&\frac{1}{\Omega}\sin(t\Omega) \\
    -\Omega\sin(t\Omega)& \cos(t\Omega)
    \end{array}\!\right)\!
    \right)\\
$%
the evolution operators for the first and the second half cycle. From this, the
main stable region can be found for $|\Omega T|\leq 3.75$. Narrow higher-order
stable regions also exist but will further be neglected. Hence there exists a
sharp drive-frequency threshold for every Stark shift, above which trapping is
possible, irrespective of the sign of the Stark shift and independent of the
initial conditions.

Note that the finite extension of our trap, $|x|\le x_{max}$, limits the trap
depth. This can be estimated by averaging over the micromotion, which is
assumed to be smaller and faster than the macromotion. This results in a trap
depth $m \pi^2 x_{\rm{max}}^2\Omega^4/(24 \omega^2)$, typically of the order of
10\,mK ~\cite{Ghosh}. For a more realistic estimate, a detailed analysis is
made below.


It is interesting to compare molecule trapping with ion trapping in a Paul
trap~\cite{Ghosh}. An ion experiences the Coulomb force $\vec{F}_{\rm C}$,
which has zero divergence, $\vec{\nabla}\cdot\vec{F}_{\rm C}=0$. As a
consequence, the the magnitude of the force is the same for positive and
negative ions. For molecules with a linear Stark shift and in the harmonic
approximation, $\vec{\nabla}\cdot\vec{F}=0$, and there is an accidental
correspondence with the equations of motion of an ion in a Paul trap. It should
nevertheless be emphasized that trapping neutral molecules differs from ion
trapping in many ways. First, whereas ions are monopolar particles which can be
trapped in an oscillating quadrupole field, neutrals have permanent or induced
dipoles which can be trapped in an oscillating dipole field. Second, in
contrast to ions, the force on a molecule in an electric field depends
dramatically on the internal state of the molecule. Third, as is obvious from
Fig.~\ref{ACidea}, the harmonic approximation is only valid in a small region
around the center. In contrast to ions, which have vanishingly little kinetic
energy compared to the $10^5\,$K trap depth, the temperature of the molecules
is likely to be comparable to the trap depth. Hence, molecules explore the full
volume of the shallow trap, including the outer regions. These outer regions
are of paramount importance in our experiment. It is here that the harmonic
approximation breaks down. The molecule-field interaction even allows
$\vec{\nabla}\cdot\vec{F}\not=0$. This leads to different behavior for HFS and
LFS away from the center, where the time-average of the force in configurations
1) and 2) is not zero. In particular, the regions near the electrodes are net
repulsive for LFS and net attractive for HFS.

\begin{figure}[htb]\begin{center}
\epsfig{file=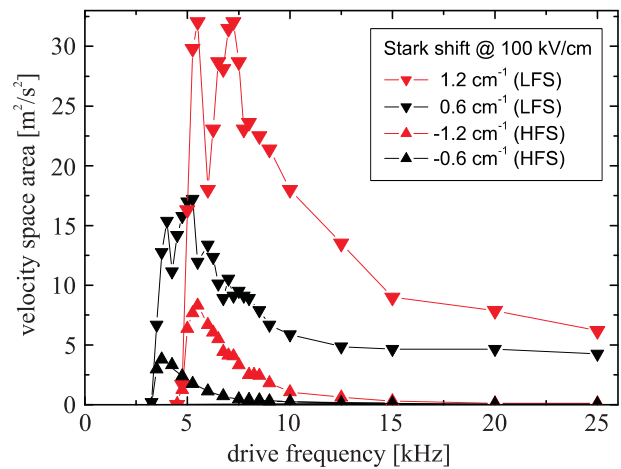,width=0.45 \textwidth}%
\caption{Area of guided molecules in transverse-velocity space obtained from a
two-dimensional Monte-Carlo simulation. Curves for LFS
($\blacktriangledown,\textcolor{red}{\blacktriangledown} $) and HFS
($\blacktriangle,\textcolor{red}{\blacktriangle} $) \NDDD\ molecules with the
given Stark shifts are displayed.
}%
\label{MathResult}
\end{center}\end{figure}

To take all these effects into account, we have performed a two-dimensional
Monte Carlo simulation. In the simulation, point-like particles are injected
on-axis in a random phase of the driving field. The input velocities $v_x$ and
$v_y$ are varied. The particles are propagated under the influence of the
periodically poled Stark force for 10\,ms, a typical time in the experiment.
They are considered lost if they leave a radius $r_{max}=1.25\,$mm. The amount
of trapped trajectories can be expressed as an area $A$ in the two-dimensional
$v_x v_y$-velocity space quadrant. $A$ is a good measure for the flux, if the
guidable velocities are equally present in the source. Results for four
different molecular states of \NDDD\ exhibiting linear Stark shifts with a
slope, $s$, of $\pm$\,0.6\,cm$^{-1}$/(100\,$\frac{kV}{cm}$) and
$\pm$\,1.2\,cm$^{-1}$/(100\,$\frac{kV}{cm}$) are plotted in
Fig.~\ref{MathResult}. For a molecular state with a Stark shift of
1.2\,cm$^{-1}$/(100\,$\frac{kV}{cm}$) and at 9\,kHz for our experimental
parameters, the occupied $v_x v_y$-space has almost a quarter-disc shape, with
which a trap depth of approximately 20\,mK can be associated.

As expected from the analytic analysis, there is a sharp turn-on frequency
above which guiding is possible both for LFS and HFS. For increasing
frequencies, the area $A$ reaches a maximum before decreasing again for higher
frequencies. Note that the maximum of $A$ is approximately four times larger
for LFS than for HFS, indicating that LFS can be guided more efficiently. This
is due to the anharmonicity of the trapping potential away from the center,
where stable trajectories exist for LFS with comparatively high initial
velocities. This effect also causes the substructure in the LFS curves in
Fig.~\ref{MathResult}. For even higher frequencies, the amplitude of the
micromotion becomes very small and the force on the molecules approaches the
time-average of the force in configurations 1) and 2). This time-averaged force
is small near the center but substantial and repulsive near the electrodes.
Stable orbits therefore exist in the high-frequency limit for LFS. The stable
orbits occupy disjunct areas in velocity space, of which the area, $A$,
saturates. The notion of a trap depth becomes difficult in this limit. The
behavior of HFS, however, is completely different: With the disappearing
micromotion, there are no stable trajectories and $A$ vanishes in the limit of
high drive frequency.



Obviously, the motion in the full anharmonic potential can be very complicated.
In fact, we have found numerical evidence for chaotic motion for some initial
conditions. But despite of the complicated dynamics, LFS and HFS are
simultaneously trapped and there is a broad overlap between the trapping
regions for different Stark shifts at the same frequencies, as evident from
Fig.~\ref{MathResult}.


\begin{figure}[htb]\begin{center}
\epsfig{file=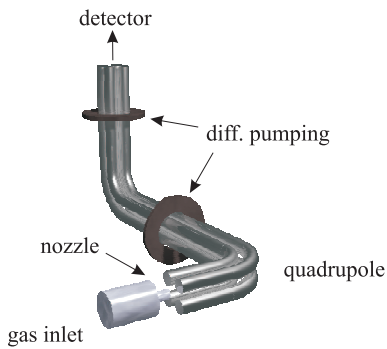,width=0.35 \textwidth}%
\caption{Schematic of the experimental setup. \NDDD\ molecules emerging from
the nozzle are injected into the guide. The slowest molecules are kept within
the guide and after passing two differential pumping stages they enter a UHV
chamber where they are detected with a mass spectrometer.
The fast molecules are pumped away.}%
\label{setup}
\end{center}\end{figure}

Our experimental demonstration of AC guiding uses an effusive source of thermal
polar \NDDD\ molecules, see the setup shown in
Fig.~\ref{setup}~\cite{Rangwala}. The double-bent guide consists of four wires
which pass through two vacuum chambers with two differential pumping sections
before ending in a UHV detection chamber. The guide has a length of 500\,mm and
is made of 2\,mm diameter stainless steel rods, with a 1\,mm gap between
neighboring rods. The radii of curvature of the two bends are 25\,mm and the
rods are built around the 0.8\,mm inner diameter ceramic nozzle, constituting
the effusive source. Typical operation pressures in the nozzle are around
0.05\,mbar in order to maintain molecular-flow conditions. Most of the
molecules are not guided and escape into the first vacuum chamber, where an
operational pressure of a few times $10^{-7}\,$ mbar is maintained. In the
detection chamber, where a pressure below $10^{-10}\,$ mbar is achieved, the
guided molecules are detected with an efficiency of about
$10^{-4}$\,counts/molecule by a quadrupole mass spectrometer (QMS)[Hiden
Analytical, HAL 301/3F] with a pulse-counting channeltron. The pulses from the
channeltron are processed in a multi-channel scaler. The ionization volume of
the QMS begins 22\,mm behind the guide. To protect the QMS from the high
electric fields, a metal shield with a 5\,mm diameter hole is placed 1\,cm
behind the exit of the guide. The time-varying electric fields are generated by
switching high voltages with fast push-pull switches. Switching frequencies are
typically in the range of a few kHz. In each phase of the alternating field,
opposite rods carry voltages of up to $\pm\,7\,$kV. Due to the inversion
splitting of the vibrational ground state, \NDDD\ shows in good approximation a
linear Stark shift in the relevant electric field range (0-100)\,kV/cm
~\cite{Schawlow}.


\begin{figure}[htb]\begin{center}
\epsfig{file=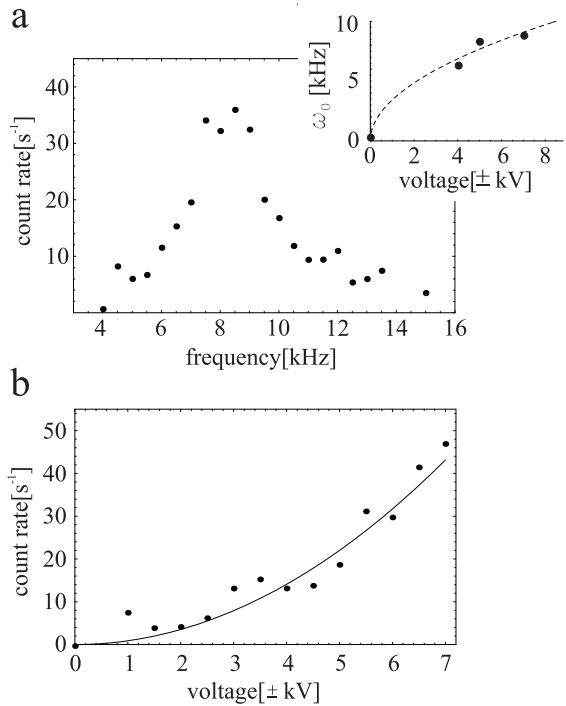,width=0.43\textwidth} \caption{(a) Molecular flux as a
function of the applied frequency for rod voltages of $\pm5\,$kV. The graph in
the inset shows the frequency of maximum flux, $\omega_{0}$, as a function of
the applied voltage. The dashed line indicates the $\omega_{0}\propto \sqrt{V}$
dependence. (b) Flux as a function of the applied voltage. For each voltage,
the frequency is adjusted for maximum flux according to (a). The line is a
quadratic fit.} \label{freqdep}
\end{center}\end{figure}

For detecting \NDDD\ molecules the QMS is set to mass 20. The influence of
other gases at this mass is negligible. The QMS does not provide
state-selective detection, which is very difficult to achieve if one considers
the variety of molecular states involved in a thermal (room-temperature)
ensemble. Our simulation, however, shows that the detection of HFS is
suppressed, because under the influence of the high electric fields most of the
HFS turn around at the exit of the guide, moving backwards on its outside, or
they are reflected back into the guide.

To discriminate the guided molecules from background gas and dark counts, the
alternating electric field is periodically applied for 200\,ms followed by a
200\,ms where no field is applied. This on-off sequence is repeated about 40000
times to obtain a good signal-to-noise ratio. From the instant when the
alternating field is switched on, the molecules can propagate along the guide
and the signal rises. After about 25\,ms, the signal has reached 50\% of the
maximum increase, corresponding to velocities around 20\,m/s or about 0.5\,K.
When the alternating field is switched off the flux instantaneously falls off.
The flux of guided molecules is determined by subtracting the background count
rate from that during the last 50\,ms of the 200\,ms-long interval where the
field is applied. The same measurement without injecting gas yields no signal.

We now investigate the frequency dependence of the system. For electrode
voltages of $\pm$5\,kV, the switching frequency is varied between (4-15)\,kHz,
and the flux of guided molecules is recorded in frequency steps of 500\,Hz. The
experimental data in Fig.~\ref{freqdep}a show that the signal amplitude reaches
a pronounced maximum at a frequency of $\approx$8\,kHz.
No discernible signal is observed for frequencies below 4\,kHz. This behavior
can be understood from the simple analytic theory above, acknowledging the
thermal ensemble of molecular states. For every molecular state there exists a
cut-off frequency corresponding to the associated Stark shift below which no
guiding is possible. This frequency increases with the Stark shift. For
example, the cut-off frequencies for molecules with Stark shifts of
0.6\,cm$^{-1}$ and 1.2\,cm$^{-1}$ at an electric field strength of 100\,kV/cm
in our guide are 3.3\,kHz and 4.7\,kHz, respectively. For small frequencies
stable trajectories exist only for molecules with small Stark shifts. These
molecules are unlikely to be guided as the potential depth is too small
compared to their kinetic energy. Higher frequencies allow stable trajectories
for molecules with higher Stark shifts and so the number of guidable states
increases. Furthermore the guiding efficiency for molecules with higher Stark
shifts is much higher. Both effects cause the rise in the count rate for
frequencies below 8\,kHz. For higher frequencies the count rate decreases
because the guided flux for every molecular state depends on the depth of the
potential which drops off with $\omega^{-2}$.

Our conclusions are supported by two additional measurements. First, the
dependence between the applied voltage and the frequency $\omega_{0}$ for which
optimum guiding exists is analyzed. In the harmonic approximation, the stable
region is given by $|\Omega T|\leq 3.75$ with $\Omega\propto \sqrt{\beta}$. As
the curvature $\beta$ scales linearly with the applied voltage $V$, we obtain
$\omega_{0}\propto \sqrt{V}$. Experimental data are shown together with the
expected square root dependence in the inset of Fig.~\ref{freqdep}a.
Unfortunately, the measurements cover only a small voltage interval, because
for small voltages no guiding signal could be obtained, whereas sparks occurred
for higher voltages. Nevertheless, the measured data agree with the calculated
scaling. This allows one to choose the optimum switching frequency for a given
voltage. In a second measurement, the flux is measured as a function of the
applied voltage, using the optimum frequency for each voltage. The experimental
data in Fig.~\ref{freqdep}b show a quadratic rise in flux with increasing
voltage. This is expected for \NDDD, where most states have a linear Stark
shift~\cite{Rangwala}.

To summarize, two-dimensional trapping of neutral molecules with alternating
electric fields is demonstrated experimentally. Our result proves that
molecules with transverse and longitudinal temperatures of 20\,mK and 0.5\,K,
respectively, can be filtered out of a room temperature effusive source
efficiently. Our simulation shows that both HFS and LFS are guided, whereas the
low-field seekers are guided and detected with a higher efficiency. The guiding
efficiency for both classes will further increase if the injected molecules are
precooled with cryogenic methods. An important aspect of our guide is that it
is suited to trap laser-cooled atoms as well~\cite{Shimizu92}. Two-dimensional
trapping of both atoms and molecules
should therefore be possible. 
Finally, an extension of our technique to trap sufficiently slow molecules in
three dimensions~\cite{Shimizu92} should be feasible.


\end{document}